\documentclass[prd,nofootinbib,tightenlines,showkeys,showpacs]{revtex4}
\usepackage{amsfonts}
\usepackage{amssymb}
\usepackage{amsmath}
\usepackage{indentfirst}
\usepackage[latin1]{inputenc}

\begin{document}
\title{$\kappa$-deformed Dirac Equation}

\author{E. Harikumar \footnote{harisp@uohyd.ernet.in}, M. Sivakumar\footnote{mssp@uohyd.ernet.in}, and N. Srinivas}

\affiliation{School of Physics, University of Hyderabad, Hyderabad 500046, India}

\vspace*{1cm}

\begin{abstract}

We construct a Dirac equation in $\kappa$-Minkowski spacetime and analyse its implications.
This $\kappa$-deformed Dirac equation is expanded as a power series involving  derivatives with respect to commutative coordinates and the deformation parameter, $a$. We show that the $\kappa$-deformation breaks the charge conjugation invariance but preserves parity and time reversal. We then study how the Hydrogen atom spectrum is modified due to the $\kappa$-deformation, applying perturbation theory. Using this, we obtain bounds on the deformation parameter $a$, which are few orders higher than the Planck length.  We also show that the effects of deformation on the spectrum are distinct from that of Moyal deformation and generalized uncertainty principle.
\end{abstract} 



\maketitle

\section{ Introduction}

The notion of modification or deformation of the Poincare symmetry due to quantum gravity effects is being studied intensively in recent times. Different approaches to understand the microscopic theory of gravity  invariably introduces  the notion of a fundamental length scale. Noncommutative geometry provides a natural way of incorporating this fundamental length scale\cite{connes,sergiod,sergiod1}. With this motivation, considerable amount of efforts have been directed to understand various types of noncommutative spacetimes as well as in analysing physical implications of these spacetimes.

The Moyal spacetime whose coordinates ${\hat X}^\mu$ satisfying $[{\hat X}^\mu, {\hat X}^\nu]=i\theta^{\mu\nu}$, where $\theta^{\mu\nu}$ is a fixed, anti-symmetric tensor is one prototype of noncommutative spacetime that has been studied extensively in recent times\cite{rev1,rev2}. Many physical models have been generalized to Moyal spacetime and the effect of noncommutative parameter $\theta^{\mu\nu}$ have been analyzed\cite{jab,jab1,jab2,jab3, ncgrav, ncgrav1, ncgrav2, ncgrav3}.

The usual notion of Lorentz symmetry is lost in the noncommutative spacetime. But it was shown that using a Hopf algebra approach, one can retain this invariance\cite{chaichian, wess1,wess1a,majid}. This allows to take over the usual notion of field quanta labeled by the Casimirs of Poincare algebra to noncommutative field theories. But this Hopf algebra approach results novel features in noncommutative field theories that are not shared by their commutative counterparts. Widely studied among these is the twisted statistics and its effects\cite{bals,bals1,bals2,bals3}.

Another class of noncommutative spacetime that have been studied is the one where the coordinates satisfy a Lie algebra type commutation relation, well known example being the fuzzy sphere\cite{madore,balbook,balbook1,balbook2,balbook3}. Kappa-deformed Minkowski spacetime is another well known example of this type spacetime. The coordinates of $\kappa$-Minkowski spacetime satisfy
\begin{equation}
 [{\hat x}^i, {\hat x}^j]=0, [{\hat x}^0,{\hat x}^i]=ia x^i , 
(a=\frac{1}{\kappa}).\label{kcom}
\end{equation} 
$\kappa$ Minkowski spacetime naturally appears as the spacetime associated with the low energy limit of certain quantum gravity models\cite{lukierski,lukierski1,lukierski2,lukierski3}. The symmetry algebra of this spacetime is the $\kappa$ Poincar\'e algebra and it is known to be related to deformed special relativity (DSR) \cite{dsr} (a modified relativity principle having a fundamental parameter of length dimension  in addition to the velocity of light). This led to the study of $\kappa$-space-time and physics on $\kappa$-space-time in recent times\cite{wess,
wess2,wess3,wess4,glikman,glikman1,glikman2,glikman3,mdljlm,mdljlm1,sm,sm1,sm2,sm3,sm4,sm5,us,us1,us2,us3,us4,pos}, bringing out its many interesting aspects.

Many different field theory models have been constructed on $\kappa$-space-time in recent times and various aspects of these models have been analyzed\cite{lukierski,lukierski1,lukierski2,lukierski3,wess,wess2,wess3,wess4,glikman,glikman1,glikman2,glikman3,mdljlm,mdljlm1,sm,sm1,sm2,sm3,sm4,sm5}. It is known that there are many different proposals for Klein-Gordon equation in $\kappa$-space-time \cite{lukierski,lukierski1,lukierski2,lukierski3, wess,wess2,wess3,wess4,sm,sm1,sm2,sm3,sm4,sm5}, all satisfying the criterion of invariance under $\kappa$-Poincare algebra. There are some works where $U(1)$ gauge theory on  $\kappa$-spacetime had been constructed\cite{mdljlm,mdljlm1}. The $U(1)$ theory in $\kappa$-spacetime is constructed in terms of commutative fields using $*-$product and Seiberg-Witten map, up to first order in the deformation parameter.

In this paper, we construct a Dirac equation associated with $\kappa$-spacetime and analyse some of its implications. In \cite{lukierski,lukierski1,lukierski2,lukierski3},  a Dirac equation  was constructed by demanding that its square should be the $\kappa$-deformed Klein-Gordon equation. But the Dirac equation obtained there was not invariant under spin-half representation of $\kappa$-Poincare algebra and it was shown how to avoid this \cite{dirac3}. But unlike in the commutative case, the square of this $\kappa$ deformed Dirac equation, obtained in \cite{dirac3} was related to the $\kappa$-deformed Pauli-Lubanski vector.  In \cite{dirac1,dirac1a} a Dirac equation consistent with deformed special relativity was obtained in the momentum space. This was shown to emerge naturally in the $\kappa$-spacetime, with a specific choice for the differential calculus defined on $\kappa$-spacetime. Using a different approach,Dirac equation satisfied by the particle governed by deformed special relativity laws was derived in  \cite{dirac2}.  It was shown that this Dirac equation is same as that on $\kappa$-spacetime. But, the anti-particle equation consistent with deformed special relativity was shown to be different from the same on $\kappa$-spacetime. Some other approaches towards the construction of Dirac equation on $\kappa$-spacetime were attempted in \cite{dirac4,dirac4a,dirac5,dirac5a}.

Here, we construct and study Dirac equation relevant for $\kappa$ spacetime using an alternate approach. In our approach, the $\kappa$-deformed Dirac equation is constructed  using the derivative operators which transform like vectors under the undeformed $\kappa$  Poincar\'e algebra (See Eqn.(\ref{diracder1}) below).  This guarantees that the Dirac equation is invariant under the spin-half representation of the undeformed $\kappa$-Poincare algebra. The generators of this undeformed $\kappa$  Poincar\'e algebra can be re-expressed in terms of the commutative coordinates and their derivatives \cite{sm,sm1,sm2,sm3,sm4,sm5}. Thus, the $\kappa$-Dirac equation we obtain is written in terms of the derivatives defined in commutative spacetime.  Using this Dirac equation, we study the effects of the $\kappa$ deformed noncommutative spacetime on Hydrogen atom spectrum. 

This paper is organized  as follows. In the next section, we provide a brief summary of the undeformed Poincare algebra and the $\kappa$-deformed Klein-Gordon equation ( in momentum space) essential for our purpose.  In section III, we first construct the $\kappa$-Dirac equation. Then we discuss the status of discrete symmetry of Dirac equation. We show that the charge conjugation is {\it not} a symmetry of our $\kappa$-Dirac equation, but  is invariant under parity and time reversal. In section IV, we discuss the relativistic Hydrogen atom in the $\kappa$-spacetime. In these discussions, we keep only terms up to first order in the deformation parameter $a(=\kappa^{-1})$. By treating the $\kappa$-dependent terms as perturbation, we find the  changes in the spectrum of Hydrogen atom; in particular, we calculate the shift in the $1S_{\frac{1}{2}}$ and $2S_{\frac{1}{2}}$ levels. Comparing this with the 
experimental results  on the energy difference between $1S$ and $2S$ levels of Hydrogen atom spectrum, we obtain  an upper bound $a<10^{-29}m$.  In section V, we study the non-relativistic limit of the $\kappa$-Dirac equation. Further, using this, we analyse the shift in the hydrogen atom spectrum (perturbatively).  The $1S-2S$ calculations set $a<10^{-26}m$ while $2P_{\frac{1}{2}}-2S_{\frac{1}{2}}$ leads to a bound $a<10^{-25}m$. The shift in the ground state energy of non-relativistic spectrum leads to the strongest bound $a<10^{-19}m$ using the current experimental data on 
Rydberg energy.  Our conclusions are given in Section VI.

\section{ $\kappa$-Poincare Algebra}

It was shown in \cite{sm,sm1,sm2,sm3,sm4,sm5,us,us1,us2,us3,us4} that the undeformed $\kappa$-Poincar\'e algebra is suitable to analyze the symmetries of $\kappa$ Minkowski spacetime. The generators of the undeformed $\kappa$-Poincar\'e algebra  are $M_{\mu\nu}$ and the modified derivative operators called the (Dirac) derivatives $D_\mu$ which transform as a vector under $M_{\mu\nu}$. The generators of this underlying symmetry algebra can be expressed in terms of the operators defined in the commutative spacetime, using a class of map between the coordinates and their derivatives of $\kappa$-Minkowski spacetime and those of commutative spacetime\cite{sm,sm1,sm2,sm3,sm4,sm5}. These maps are characterized by a function $\varphi$\cite{sm,sm1,sm2,sm3,sm4,sm5}.

The undeformed $\kappa$ -Poincar\'e algebra is defined through the relations
\begin{eqnarray}
&[M_{\mu\nu}, D_\lambda]=\eta_{\nu\lambda}D_\mu-\eta_{\mu\lambda}D_\nu,~ [D_\mu,D_\nu]=0,&\\
&[M_{\mu\nu}, M_{\lambda\rho } ]=\eta_{\mu\rho }M_{\nu\lambda } +
\eta_{\nu\lambda}M_{\mu\rho}
  - \eta_{\nu\rho }M_{\mu\lambda } -   \eta_{\mu\lambda}M_{\nu\rho},\label{diracder1}
\end{eqnarray}
which were obtained in \cite{sm,sm1,sm2,sm3,sm4,sm5}. Here $\eta_{\mu\nu}={\rm diag}(-1,1,1,1)$. $D_\mu D^\mu$, the Casimir of this algebra which can be expressed in terms of a $\square$ operator as 
$D_\mu D^\mu=\square(1-\frac{a^2}{4}\square)$\cite{sm,sm1,sm2,sm3,sm4,sm5,us,us1,us2,us3,us4,wess,wess2,wess3,wess4}
where the $\square$ operators are defined by $[M_{\mu\nu},\square]=0,$ $~[\square, {\hat x}_\mu]=2D_\mu.$

The co-products are defined as
\begin{eqnarray}
\Delta(D_\mu)&=& D_\mu \otimes I + I \otimes D_\mu +ia_\mu (D^\alpha Z)\otimes D_\alpha 
-\frac{a_\mu}{2}\Box Z\otimes a_\alpha D^\alpha\\
\Delta(M_{\mu\nu}) &=& M_{\mu\nu}\otimes I+I\otimes M_{\mu\nu} +
ia_\mu(D^\alpha-\frac{ia^\alpha}{2}\Box)Z\otimes M_{\alpha\nu} -ia_\nu(D^\alpha-\frac{ia^\alpha}{2}\Box)Z\otimes M_{\alpha\mu},
\end{eqnarray}
where $Z^{-1}=ia D_0 +\sqrt{1+a^2 D_\alpha D^\alpha}$.

For arbitrary realizations characterized by $\varphi$, these Dirac derivatives and $\square$ are explicitly given as
\begin{eqnarray}
&D_i=\partial_i\frac{e^{-A}}{\varphi},~~
D_0=\partial_0\frac{sinhA}{A}+ia{\vec \nabla}^2\frac{e^{-A}}{2\varphi^2},\label{d2}&\\
&\square ={\vec \nabla}^2\frac{e^{-A}}{\varphi^2}+2\partial_{0}^2 \frac{(1-coshA)}{{A^2}} \label{box}&
\end{eqnarray}
where ${\vec \nabla}^2=\partial_i\partial_i$ and $A=-ia\partial_0$. Note that $\partial_i$ and $\partial_0$ are the derivatives corresponding to the commutative spacetime coordinates. It is clear that the Casimir, $D_\mu D^\mu$ reduces to the usual relativistic dispersion relation in limit $a\to 0$.

Generalized Klein-Gordon equation using the Casimir on $\kappa$-space is  written as 
$\square(1-\frac{a^2}{4}\square)\Phi(x)-m^2\Phi(x)=0$ \cite{sm,sm1,sm2,sm3,sm4,sm5,us,us1,us2,us3,us4}. This leads to the deformed dispersion relation
\begin{equation}
\frac{a^2}{4}Sinh^2(\frac{ap_0}{2})-{\vec p}\cdot{\vec p}\frac{e^{-ap_0}}{\varphi^2(ap_0)} -m^2-\frac{a^2}{4}\left[\frac{a^2}{4}Sinh^2(\frac{ap_0}{2})-{\vec p}\cdot{\vec p}\frac{e^{-ap_0}}{\varphi^2(ap_0)}\right]^2=0.\label{dis}
\end{equation}
where $p_0=i\partial_0$ and $p_i=-i\partial_i$.

Since the Casimir as well as the $\square$ operator have the same $a\to 0$ limit, the requirement of correct Klein-Gordon equation in the commutative limit does not rule out other possible generalizations in the $\kappa$-space \cite{us,us1,us2,us3,us4}.

In this study, we take $\varphi=e^{-A}$. With this  choice\footnote{Note that, for $\psi=1$ realization we are interested in here,  we have $\varphi^\prime(0)=1-\gamma(A)$ which is always a number.  Now, for an arbitrary choice of $\varphi$ (satisfying all the consistency conditions), using Taylor series, we have, up to first order in $a$, $\varphi(A)=1+ia\varphi^\prime(0)\partial_0 $. This differs from the choice $\varphi=e^{-A}=1+ia\partial_0$ only by a numerical factor and hence the general conclusions we arrive at here will be valid, independent of our  choice for $\varphi(A).$},  the dispersion relation is same as that of $\kappa$  Poincar\'e algebra in bi-crossproduct basis \cite{majid1,majid1a} which is relevant for Dirac equation compatible with doubly special relativity \cite{dirac1,dirac1a}.

\section { $\kappa$-Dirac Equation and PTC symmetry}

In this section, we construct $\kappa$-Dirac equation such that its square gives the $\kappa$-deformed Klein-Gordon equation given above in Eqn.(\ref{dis}). Then we study the symmetry ( or the lack of it) of this Dirac equation under discrete transformations of parity, time reversal and charge conjugation.

In terms of the Dirac derivatives in 
Eqn.(\ref{d2}), above $\kappa$-Dirac equation can be written as 
\begin{equation}
(\gamma^0 D_0 +\gamma^i D_i+\frac{mc}{\hbar})\Psi=0\label{diraceqn}.
\end{equation}
Square of this equation reproduces the Klein-Gordon equation given in Eqn.(\ref{dis}), as required. 

We emphasis that the above $\kappa$-Dirac equation is defined in the commutative spacetime and the corresponding  $\gamma$ matrices are independent of the deformation parameter $a$. We take these matrices as $\gamma^0=-i\beta,~~\gamma^i=-i\beta\alpha^i$.

\subsection{Parity} Under parity, we have$
P:x\rightarrow -x,$ $~ P:t\rightarrow t,$ $~P:D_i\rightarrow -D_i,~$ $P:D_0\rightarrow D_0.$
Note that $D_i$ and $D_0$ transform respectively as pseudo vector and scalar, under parity. Thus, under parity, the $\kappa$-Dirac equation becomes
\begin{equation}
 (\gamma^0 D_0 -\gamma^i D_i+\frac{{mc}}{\hbar})\Psi(-x,t)=0\label{pdiraceqn}.
\end{equation}
and it is easy that ${\cal P}\Psi=\gamma_0 P\Psi(x,t)$ is the solution of the Dirac equation given in Eqn.(\ref{diraceqn}). Hence parity is a symmetry of the above $\kappa$-Dirac equation.

\subsection {Time reversal}

 Under time reversal, we have 
${\cal T}:x\rightarrow x,~$ ${\cal T}:t\rightarrow -t,$ $~{\cal T}:D_i\rightarrow D_i,~$ 
$ {\cal T}:D_0\rightarrow {\tilde{ D}}_0$
where
\begin{equation}
 {\tilde {D}}_0=-\frac{i}{a} sinh(-ia\partial_0)+\frac{ia}{2} \nabla^2 e^{+ia\partial_0}\label{td0}.
\end{equation}
We note that ${\tilde{D}}_{0}^*=-D_0$.
After re-expressing the Dirac equation in Eqn.(\ref{diraceqn}) as
\begin{equation}
 i\hbar D_0 \Psi=\left[-i\hbar\alpha\cdot D +\beta mc\right]\Psi\equiv H\Psi\label{deqn}
\end{equation}
we see easily that under the time reversal, the Dirac equation becomes
\begin{equation}
 i\hbar{\tilde{D}}_0\Psi(x,-t)=H\Psi(x,-t).
\end{equation}
Taking complex conjugate, we get
\begin{equation}
i\hbar D_0\Psi^*(x,-t)= H^*\Psi^*(x,-t)
\end{equation}
from which we see that ${\cal T}\Psi^*(x,t)$ satisfies the Dirac equation in Eqn.(\ref{deqn}). Here the time reversal operator is given as ${\cal T}=-i\alpha_1\alpha_3$.
As in the case of parity, time reversal is also a symmetry of the $\kappa$-Dirac equation. It is interesting to note that both parity and time reversal operators are the same as that in the commutative case.

\subsection {Charge Conjugation}

To study the charge conjugation, we introduce minimal coupling of electromagnetic fields in Eqn.(\ref{deqn}) and obtain the Dirac equation for a charged particle interacting with external electromagnetic field . The minimal prescription is to replace $p_\mu (=-i\hbar\partial_\mu)$ with $ p_\mu-eA_\mu$ where $e$ is the electric charge of the particle. With this, the Dirac equation becomes
\begin{equation}
 i\hbar\left( \frac{i}{a}sinh[a(p_0-eA_0)]-\frac{ia}{2\hbar^2}({\vec p}-e{\vec A})^2 
e^{a(p_0-eA_0)}\right)\Psi=\left({\vec\alpha}\cdot({\vec p}-e{\vec A}) +\beta mc\right)\Psi.
\label{deq3}
\end{equation}
To get the equation for the anti-particle, we change $e$ to $-e$ in the above equation and take the complex conjugate, leading to
\begin{equation}
i\hbar\left( \frac{i}{a}sinh[a(p_0-eA_0)]+\frac{ia}{2\hbar^2}({\vec p}-e{\vec A})^2 e^{-a(p_0-eA_0)}\right)\Psi^*=\left({\vec\alpha}^*\cdot({\vec p}-e{\vec A}) -\beta mc\right)\Psi^*.
\end{equation}

We note that a matrix $C=i\beta\alpha_2$ satisfying $C\alpha^* C^{-1}=\alpha,$ $~C\beta^*C^{-1}=-\beta$ will map the RHS of the above equation to that of Eqn.(\ref{deq3}) by a similarity transformation.

But note that, the LHS of the above equation is identity matrix in the spinor space
and the same similarity transformation will not change the sign of the second term as well as the sign of the exponential appearing in that term. This shows that the Dirac equation for anti-particle is different from that of the particle. This feature of Dirac equation on $\kappa$ spacetime was noticed in \cite{dirac2} using a different approach. Thus we see here that charge conjugation is not a symmetry of $\kappa$ Dirac equation.

We note that if we allow the sign of the deformation parameter $a$ to change under the operation of charge conjugation we still can have the same Dirac equation for both particle and anti-particle. But then we can no longer consider $a$ to be a fundamental constant.

\section{ $\kappa$-Dirac equation and Spectrum of Hydrogen atom}

In this section we study the modifications in the energy spectrum of relativistic Hydrogen atom due to the $\kappa$-deformation. This is done by keeping terms up to first order in the deformation parameter and studying its effect using first order perturbation theory. Using the shift in $1S_{\frac{1}{2}}$ and $2S_{\frac{1}{2}}$ calculated, and from the experimental values, we set a bound on the deformation parameter.

With applying perturbation treatment in mind, we  expand $D_0$ and $D_i$ appearing in Eqn.(\ref{diraceqn}) in powers of $a$ and 
obtain the $\kappa$ Dirac equation, valid  up to first order in $a$. To this order, the $\kappa$ Dirac equation becomes
\begin{equation}
 i\hbar \partial_t\Psi=-i\hbar c{\vec \alpha}\cdot{\vec \nabla}\Psi +mc^2\beta\Psi
+\frac{ac\hbar}{2} {\vec\nabla}^2 \Psi.\label{moddir}
\end{equation} 
Note here that the only $a$ dependent term, i.e., $\nabla^2$ is an identity operator in the spinor space. Hence the total angular momentum is still a good quantum number, i.e., $J=L+S$ commutes with the above Hamiltonian. 

This $a$ dependent term in the Hamiltonian leads to an additional term in the Heisenberg equation for the coordinates $x_i$. From Eqn. (\ref{moddir}) it is clear this additional term in the expression for velocity is linear in momentum. Therefore the Zitterbewegung is not affected by $a$ dependent corrections (up to first order).

The Dirac equation for Hydrogen atom is now
\begin{equation}
 i\hbar \partial_t\Psi=\left[-i\hbar c{\vec \alpha}\cdot{\vec \nabla} +mc^2\beta +V(r)
+\frac{ac\hbar}{2} {\vec\nabla}^2 
\right]\Psi\equiv(H_0+H_1)\Psi,
\end{equation}
where $V(r)=-\frac{Ze^2}{4\pi\epsilon_0 r}$ is the Coulomb potential.\footnote{ It was shown in \cite{ehakk}, that the change in the $\frac{1}{r}$ potential due to $\kappa$ deformation, up to first order in the deformation parameter $a$ is a total time derivative. Also this additional term do not contribute to the Hamilton's equation}.  Note that the perturbing Hamiltonian  $H_1=a\frac{c\hbar}{2}{\vec \nabla}^2$. Thus the first order perturbation to energy spectrum is
\begin{equation}
\Delta E=<\psi|H_1 |\psi>.
\end{equation}
This can be calculated using the well known expression for the relativistic Dirac wave function \cite{paulstrange} 
\begin{equation}
 \psi= \left(\begin{array}{c}
g_k(r) \chi_{k}^{mj}(\hat r)\cr
i f_k(r) \chi_{-k}^{mj}(\hat r)
\cr
\end{array}\right)
\end{equation}
where $\chi_{k}^{mj}$ are the spin-angular functions, $g_k(r)$ and $f_k(r)$ are radial wavefunctions. Using the Laplacian operator in spherical polar coordinates, we  get the first order correction as
\begin{eqnarray}
 \Delta E&=&\int d^3 r g_{k}(r)\chi_{k}^{{m_j}\dagger}
(\hat r)\left(\frac{1}{r^2}\frac{\partial}{\partial r} (r^2 \frac{\partial}{\partial r})-\frac{L^2}{\hbar^2r^2}\right) g_{k^\prime}(r)\chi_{k^\prime}^{m_{j}^\prime}(\hat r)\nonumber\\
&+&\int d^3 r f_{k}(r)\chi_{k}^{{m_j}\dagger}
(\hat r)\left(\frac{1}{r^2}\frac{\partial}{\partial r} (r^2 \frac{\partial}{\partial r})-\frac{L^2}{\hbar^2r^2}\right) f_{k^\prime}(r)\chi_{k^\prime}^{m_{j}^\prime}(\hat r)\label{e1}
\end{eqnarray}

The orthogonality relations of $\chi_{k}^{m_j}$ and the identity
\begin{equation}
 \int d\Omega \chi_{k^\prime}^{{m_{j}^\prime}}(\hat r) L^2 
\chi_{k}^{{m_j}}(\hat r)=l(l+1)\hbar^2 \delta_{k^\prime k} \delta_{m_{j}^\prime m_{j}},
\end{equation}
we calculate the first order correction to $1S_{\frac{1}{2}}$ and $2S_{\frac{1}{2}}$ states of the Hydrogen atom.

Using the expressions for $g(r)$ and $f(r)$ for $1S_{\frac{1}{2}}$ (see \cite{paulstrange}) in Eqn(\ref{e1}), we find, for $1S_{\frac{1}{2}}$,
\begin{equation}
 \Delta E_1=\frac{ac\hbar (2\lambda)^{2s+1}}{2 \Gamma(2s+1)}\int dr r^{s-1}e^{-\lambda r}
\left(2r\frac{d}{dr} +r^2\frac{d^2}{dr^2}\right) r^{s-1}e^{-\lambda r}.
\end{equation}
Here $\lambda= \frac{1}{\hbar}\sqrt{(m^2c^2- E^2 c^{-2})}$, where $E$ is the corresponding unperturbed energy  and $s=\sqrt{1-\alpha^2}$, $\alpha$ being the fine structure constant. Approximating $s$ to unity, we find
\begin{equation}
  \Delta E_1 =-0.10256 a ~J. 
\end{equation}

Now using $g(r)$ and $f(r)$ corresponding to $2S_{\frac{1}{2}}$, we find
\begin{equation}
 \Delta E_2= -236.32 a ~J.
\end{equation}

Using these, we find the ratio of the difference in the shifts of $1S$ and $2S$ levels to the energy of $1S$ level. This turns out to be
\begin{equation}
 |\frac{\Delta E_1-\Delta E_2}{E_1}|=a (2.89\times 10^{15})m^{-1}
\end{equation}

The frequency of $1S-2S$ is a now known to an accuracy of about $10^{-14}$\cite{udem,udem1}. This imposes the bound
\begin{equation}
 a 2.89\times 10^{15}< 10^{-14}m
\end{equation}
implying $a< 10^{-29}m$. This bound is above the Planck length.

As we show below, the $\kappa$ induced shift in the spectrum of Hydrogen atom persist even in the non-relativistic limit. Since the result in the NR limit is more transparent and also provide much stronger bounds on $a$, we next study the NR limit of the Dirac equation and study the non-relativistic Hydrogen atom spectrum.

\section{ Non-relativistic Hamiltonian in $\kappa$-space-time} 

Now we study the non-relativistic limit of the $\kappa$-Dirac equation obtained in section 2. Then, we use this to study the shift the spectrum of Hydrogen atom, using perturbation theory and calculate shifts of various energy levels. Using this we obtain different bounds on the deformation parameter.

The $\kappa$ Dirac equation in Eqn. (\ref{moddir}) leads to the two component equations
\begin{equation}
\left [E-mc^2+\frac{ac}{2\hbar} {\vec p}^2-V(r)\right]U=c\sigma\cdot p W,~~
\left[E+mc^2+\frac{ac}{2\hbar}{\vec p}^2-V(r)\right]W=c\sigma\cdot p U. 
\end{equation}
Splitting $E=E^\prime +mc^2$ and $mc^2>> E^\prime$ where $E^\prime$ is the NR energy and following the standard procedure, we find, in the NR limit $U$ obeys
\begin{equation}
 E^\prime U=\left[-\frac{\hbar^2}{2m}{\vec 
\nabla}^2+V+\hbar^2\frac{(E^\prime-V)}{4m^2c^2}{\vec \nabla}^2 
+\frac{1}{2m^2c^2}\frac{1}{r}\frac{dV}{dr} (L\cdot S)-\frac{\hbar^2}{4m^2c^2}\frac{dV}{dr}\frac{\partial}{\partial r} +\frac{ac\hbar}{2}{\vec \nabla}^2-\frac{a\hbar^3}{8m^2c}({\vec\nabla}^2)^2\right] U,\label{nr1}
\end{equation}
which is in the form $Eu=(H_0+H_1+H_a)u$. Here $H_1$ is the well known correction terms one finds in the NR limit of Dirac Hamiltonian for $H$-atom and $H_a$ comprises of the last two terms induced by the $\kappa$ noncommutativity.
Note here that the usual kinetic term as well as the relativistic correction term receive $a$ dependent corrections and this lead to change in the coefficients of these two terms. But neither the spin-orbit coupling term nor Darwin term are modified by the $\kappa$ noncommutativity parameterized by $a$. Also, no new type of terms (which were not present in $H_0$ or $H_1$) are generated by $\kappa$ deformation.

The perturbative effect of $H_1$  on the spectrum and eigenfunctions of $H_0$ are well known \cite{bhbcjj}. The energy eigenvalue of $H_0+H_1$ is 
\begin{equation}
 E_n=E_{n}^0+E_{n}^0\frac{(Z\alpha)^2}{n^2}~(\frac{n}{j+\frac{1}{2}}-\frac{3}{4})
\end{equation}
where $E_{n}^0=-\frac{1}{2}mc^2\frac{(Z\alpha)^2}{n^2}, n=1,2,....$ and $\alpha$ is the fine structure constant.

The first order correction to energy eigenvalue due to $\frac{ac\hbar}{2}{\vec \nabla}^2=-\frac{ac}{2\hbar}{\vec p}^2=-\frac{acm}{\hbar}(H_0-V(r))$
is 
\begin{equation}
 \Delta E_1=-\frac{acm}{\hbar}<nlm|(H_0-V(r))|nlm>=\frac{acm}{\hbar}E_{n}^0,~ 
n=1,2,..\label{corr1}.
\end{equation}
Similarly, the first order correction due to $-\frac{a\hbar^3}{8m^2c}({\vec \nabla}^2)^2=-\frac{a}{8m^2\hbar c}({\vec p}^2)^2=-\frac{a}{2\hbar c}(H_0-V(r))^2$ is 
\begin{equation}
 \Delta E_2= -\frac{a}{2\hbar c}<nlm|(H_0-V(r))^2|nlm>=
-\frac{amc}{\hbar}E_{n}^0\left(\frac{Z\alpha}{n}\right)^2\left[\frac{3}{4}-\frac{n}{l+\frac{1}{2}}\right].\label{corr2}
\end{equation}
Thus the the energy eigenvalue of $H$ is 
\begin{equation}
 E_n=E_{n}^0\left[1+\left(\frac{Z\alpha}{n}\right)^2\left(\frac{n}{j+\frac{1}{2}}-\frac{3}{4}\right)\right]
+E_{n}^0\frac{acm}{\hbar}
\left[1- \left(\frac{Z\alpha}{n}\right)^2\left(\frac{3}{4}-\frac{n}{l+\frac{1}{2}}\right)\right].\label{Ev}
\end{equation}
Note here that the correction due to the second term (Eqn.\ref{corr2}) explicitly breaks the degeneracy in the orbital quantum number $l$. But the $\kappa$ noncommutativity induced corrections do not lift the degeneracy in $m$. 

Using the spectrum (Eqn.\ref{Ev}), we find the ratio of energy difference between $1S$ and $2S$ levels to the energy of $1S$ level to be a $ 2.5 a\times 10^{12}$. Present experimental results \cite{udem,udem1} on this ratio leads to the upper bound on $a$. Thus we have 
\begin{equation}
 a<10^{-26}m.
\end{equation} We see here that the bound on $a$ obtained in the non-relativistic limit of the $\kappa$- modified Hydrogen atom is stronger that the corresponding bound obtained in the relativistic case.

We also see from the non-relativistic spectrum ( Eqn.\ref{Ev}) that the shift
\begin{equation}
 E_{2p_{\frac{1}{2}}}-E_{2s_{\frac{1}{2}}}=a3.76\times 10^{-6} Jm^{-1}.
\end{equation}
Since the accuracy in this measurement is $10^{-31}$,
we find the upper bound as $a<10^{-25}m$.

We get a much more stronger bound from the correction to the ground state energy of Hydrogen atom due to $H_a$. As it is clear from Eqn.(\ref{corr1}) and Eqn.(\ref{Ev}), this contribution shifts the Rydberg energy. For $n=1$, we find the ratio of this shift in Rydberg energy to the unperturbed one
\begin{equation}
 |\frac{\Delta E}{E}|=\frac{a}{2} \times 10^{11}m^{-1}.
\end{equation}
Since the accuracy in the measurement of Rydberg energy is $10^{-8}$, we find the upper bound on $a$ as
\begin{equation}
 a<10^{-19}m.
\end{equation}

\section{Discussion}

In this paper we have constructed and studied Dirac equation for $\kappa$ Minkowski spacetime, using the Dirac derivatives. Since these Dirac derivatives transform like vectors under the undeformed $\kappa$-Poincare algebra, the Dirac equation is guaranteed to be covariant under the undeformed $\kappa$-Poincare algebra. We also note that all the dependence of the deformation parameter is included in the Dirac derivatives and the gamma matrices are
independent of deformation parameter. 

The Dirac derivatives are defined in terms of commutative coordinates and their 
derivatives. Since the mapping between $\kappa$ spacetime coordinates and their commutative counterparts is characterized by  $\varphi$,  the  $\kappa$-Dirac equation will naturally include $\varphi$. Thus the dispersion relation is also characterized by the same $\varphi$. The choice of $\varphi$ we made here give the same dispersion relations that obtained in the bi-crossproduct basis of $\kappa$ Poincar\'e algebra. As pointed out in the footnote, different choices for $\varphi$ will not affect the general conclusions we have obtained.

We have shown that the charge conjugation is no more a symmetry of the $\kappa$-Dirac equation whereas parity and time reversal invariances are unaffected by the $\kappa$-deformation. These results are valid to all orders in the deformation parameter. 

We have analyzed the shift in the energy levels of Hydrogen atom due to the $\kappa$-deformation, using this Dirac equation. This was done using perturbartion theory, keeping only terms up to first order in $a$. Note that the modification in $\frac{1}{r}$ potential (up to first order in $a$) is a total time derivative term and this can be omitted\cite{ehakk}. We have also obtained the non-relativistic limit of the $\kappa$-Dirac equation and using this also, we have studied the shift in the spectrum of Hydrogen atom( valid up to order $a$). Comparing these shifts in different energy levels with the experimental data, we have obtained different bound on the deformation parameter. All these bounds shown that that deformation parameter $a$ is much higher than the Planck length.

We notice that the degeneracy in the spectrum due to $l$ is removed in our case as well as in the case modification of non-relativistic Hydrogen atom spectrum due generalized uncertainty relations\cite{FB,sd,sd1}. But unlike in the later, in the present case, we see that there is also a shift in the ground state energy due to the $p^2$ term (see $H_1$). This $p^2$ term is present in the NR limit(see Eqn.(\ref{nr1})) which is not present in the case of modification due to generalized uncertainty relation. Though the NR calculations shows uniform shift in the ground state energy, we see from the Dirac equation that this shift would be different for different levels. Thus, the shift in the ground state energy, is a signature of the $\kappa$ spacetime modification which is not shared by the modification coming from the generalized uncertainty relation. In \cite{jab} it was shown that the Moyal noncommutativity opens up new channels for transitions. In the $\kappa$ spacetime, this modification is absent and the degeneracy in $m$ quantum number is still intact as in the commutative case. This can distinguish the effect of Moyal noncommutativity from that of $\kappa$ space time.

{\bf Acknowledgement}: We thank A. K. Kapoor for useful discussions. MS and NS thank DST(India) for support through  a project.

\end{document}